\documentclass[epj]{svjour}
\usepackage{graphics}
\usepackage{graphicx,epsfig,epstopdf,longtable}
\begin{document}
\title{Experimental level densities of atomic nuclei}
\author{M.~Guttormsen\inst{1}\thanks{\emph{E-mail:} magne.guttormsen@fys.uio.no}%
\and M.~Aiche\inst{2}
\and F.L.~Bello Garrote\inst{1}
\and L.A.~Bernstein\inst{3}
\and D.L.~Bleuel\inst{3}
\and Y.~Byun\inst{4}
\and Q.~Ducasse\inst{2}
\and T.~K.~Eriksen\inst{1}
\and F.~Giacoppo\inst{1}
\and A.~G\"orgen\inst{1}
\and F.~Gunsing\inst{5}
\and T.~W.~Hagen\inst{1}
\and B.~Jurado\inst{2}
\and M.~Klintefjord\inst{1}
\and A.C.~Larsen\inst{1}
\and L.~Lebois\inst{6}
\and B.~Leniau\inst{6}
\and H.~T.~Nyhus\inst{1}
\and T.~Renstr\o m\inst{1}
\and S.~J.~Rose\inst{1}
\and E.~Sahin\inst{1}
\and S.~Siem\inst{1}
\and T.G.~Tornyi\inst{1}
\and G.M.~Tveten\inst{1}
\and A.~Voinov\inst{4}
\and M.~Wiedeking\inst{7}
\and J.~Wilson\inst{6}
}                     
\institute{Department of Physics, University of Oslo, N-0316 Oslo, Norway
\and CENBG, CNRS/IN2P3, University of Bordeaux, Chemin du Solarium B.P.~120, 33175 Gradignan, France
\and Lawrence Livermore National Laboratory, 7000 East Avenue, Livermore, CA 94550-9234, USA
\and Department of Physics and Astronomy, Ohio University, Athens, Ohio 45701, USA
\and CEA Saclay, DSM/Irfu/SPhN, F-91191 Gif-sur-Yvette Cedex, France
\and Institut de Physique Nucleaire d'Orsay, B\^{a}timent 100, 15 rue G. Glemenceau, 91406 Orsay Cedex, France
\and iThemba LABS, P.O. Box 722, 7129 Somerset West, South Africa
}
\date{Received: date / Revised version: date}
%
\abstract{
It is almost 80 years since Hans Bethe described the level density as a non-interacting gas of protons and neutrons.
In all these years, experimental data were interpreted within this picture of a fermionic gas.
However, the renewed interest of measuring level density using various techniques calls for a
revision of this description. In particular, the wealth of nuclear level densities measured
with the Oslo method favors the constant-temperature level density over the Fermi-gas picture.
From the basis of experimental data, we demonstrate that nuclei exhibit a constant-temperature level density behavior
for all mass regions and at least up to the neutron threshold.
} 
\maketitle
\section{Introduction}
\label{intro}

The level density of atomic nuclei provides important information on the heated nucleonic many-particle system. In the pioneering work of
Hans Bethe in 1936~\cite{bethe36}, the level density was described as a gas of non-interacting fermions moving in equally
spaced single-particle orbitals. Despite the scarce nuclear structure knowledge at that time, this Fermi-gas model contained
the essential components apart from the influence of the pairing force between nucleons in time-reversed orbitals. The
understanding of such Cooper pairs and the nuclear structure consequences, was first realized twenty years
later through the Bardeen-Cooper-Schrieffer (BCS) theory~\cite{BCS}.

Unfortunately, the new knowledge of the pairing force was implemented in an oversimplified way by
still assuming a gas of fermions.
The original Fermi-gas level density curve was simply shifted up in energy by $\Delta$ and $2\Delta$
for the description of odd-mass and even-even mass nuclei, respectively. Later, this energy correction was found to be too large,
and the shift was somewhat back-shifted again.
The resulting back-shifted Fermi-gas model has since then been very
popular and has been the most common description of nuclear level densities for decades~\cite{capote2009}.
These manipulations maintained the typical $\exp(2\sqrt{aE})$ Fermi-gas level density shape with excitation energy $E$.

There is a growing interest in the nuclear science community for the study of nuclear level densities. The introduction of novel
theoretical approaches and fast computers has opened the way for a microscopical description of heavier nuclei up to high
excitation energy~\cite{goriely2009}.
Thus, experimental level densities represent a basic testing ground for many-particle nuclear structure models. They
also have an increasing importance for various nuclear applications. Nuclear level densities play an essential role in the calculation of
reaction cross sections applied to astrophysical nucleosynthesis, nuclear energy production and transmutation
of nuclear waste.

\begin{figure}[]
\begin{center}
\includegraphics[width=\columnwidth]{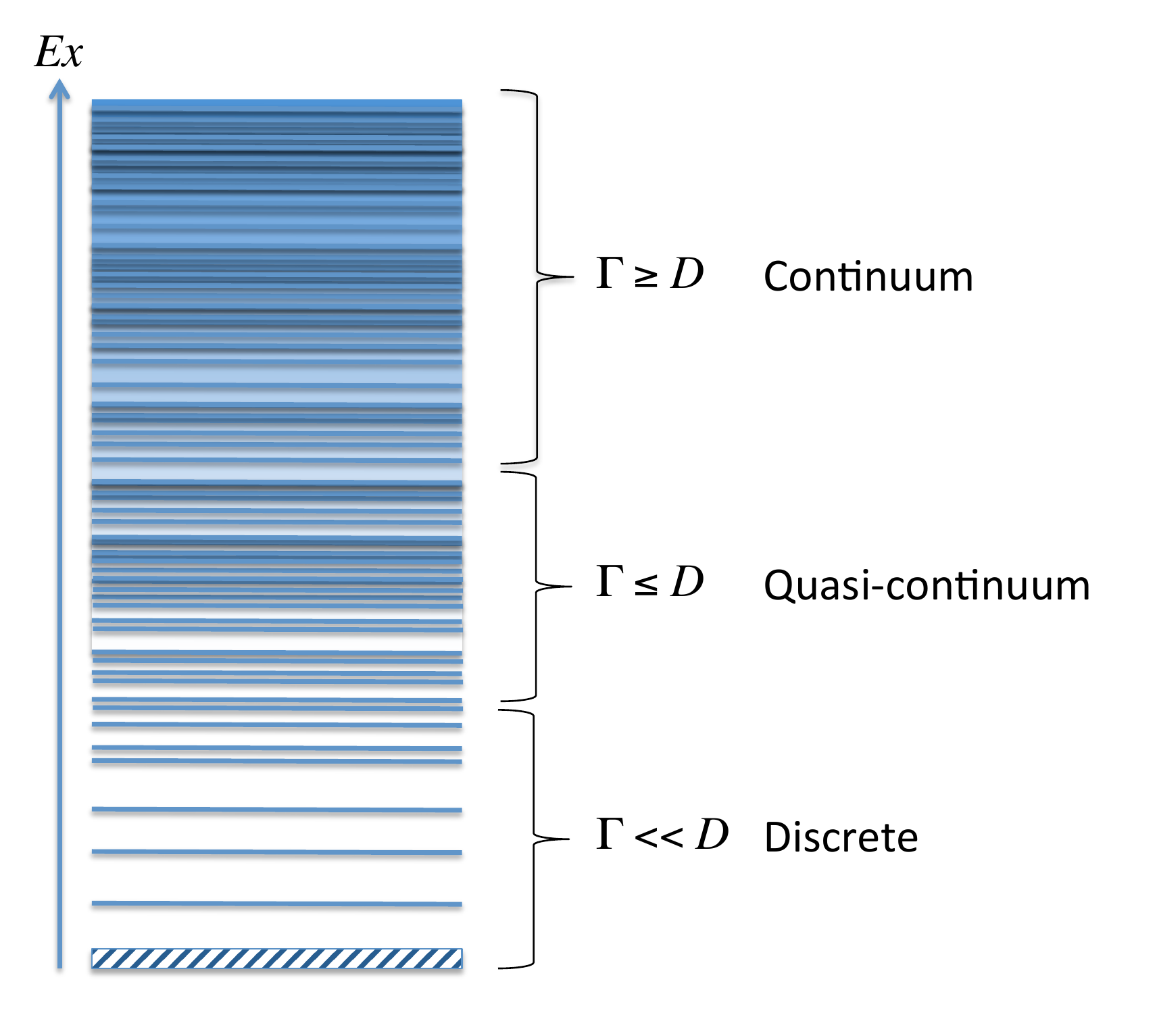}
\caption{(Color online) The energy regimes of nuclear excitations.}
\label{fig:regimes}
\end{center}
\end{figure}

The number of levels is exponentially increasing with excitation energy and typically for the rare earth region,
the level density increases by a factor of one million when going from the ground state up to the neutron binding energy.
This tremendous change has implications for the theoretical interpretations since levels start to overlap in energy at
high excitation energies, typically around 10-15 MeV. Thus, nuclear excitations are
often divided into three regimes \cite{Ericson}, classified by the average level spacings $D$ compared to the
$\gamma$-decay width $\Gamma$. These key numbers are connected to the level density and life time by $\rho=1/D$ and $\tau=\hbar/\Gamma$, respectively.
Figure~\ref{fig:regimes} illustrates the connections between the three regimes i.e.~the discrete, quasi-continuum and continuum regions.

Experimentally, the quasi-continuum
region is often defined as the region where the experimental resolution $\Delta E_{\rm res}$ prevents
the separation into individual levels. This appears when $\Delta E_{\rm res}\geq D$. One
should be aware that this definition depends on the selectivity
of the reactions used and the specific experimental detectors and conditions. In this work,
the experimental quasi-continuum region starts when $\Delta E_{\rm res}\approx D \approx 100$~keV, i.e. when the level density
exceeds $\rho\approx 10$~MeV$^{-1}$.

Today various experimental techniques are used to determine nuclear level densities. The obvious method
in the discrete region of Fig.~\ref{fig:regimes} is to count the number of
levels $\Delta N$ per excitation energy bin $\Delta E$, readily giving the level density as $\rho(E)=\Delta N/\Delta E$.
This requires that all levels are known. Making a short review on known discrete levels,
e.g.~in the compilation of Ref.~\cite{NNDC}, we
find that the data set starts to be incomplete where the first Cooper pairs are broken,
i.e. at excitation energies $E \approx 2\Delta$. This is
simply because the high density of levels makes conventional spectroscopy difficult.
A rule of thumb is that when $\rho \approx 50- 100$~MeV$^{-1}$, the number of missing levels is dramatically increasing.

Level density information at the neutron separation energy $S_n$
can be extracted from neutron capture resonance spacings $D$~\cite{RIPL3}. The
spin selection of these resonances depends on the target ground-state spin and the neutron spin transfer,
giving level densities for a narrow spin window, only. In particular, thermal neutrons with angular momentum
transfer $\ell=0$ results in a spin window of $I_{\rm target}\pm 1/2$.
At higher excitation energies the
method of Ericson fluctuations can be used~\cite{Ericson}. A well known technique~\cite{Vonach} is
to extract the level density from particle evaporation spectra.
The measurement must be carried out at backwards center-of-mass angles to avoid direct reaction contributions.
The method also requires well determined optical potential model parameters.
Furthermore,
the level density can be extracted in the two-step cascade method~\cite{Hoogenboom} from simulations
provided that the $\gamma$-ray strength function is known.
An upcoming technique, see e.g. Ref.~\cite{parity2007}, is to use high-energy light-ion reactions
like $(p, p')$, $(e, e')$ and $(^3$He, $t)$ at small angles with respect to the beam direction. These reactions select
the population of discrete levels with certain spin/parity assignments, information
that is essential in the detailed understanding of level densities.
This method requires that the spacings of the selected levels
are larger or comparable to the experimental particle resolution.

In the present work we will show level densities of several nuclei measured at the Oslo Cyclotron Laboratory (OCL).
The technique used, the Oslo method~\cite{Schiller00,Lars11}, is unique in the sense that it
allows for a simultaneous determination of the level density and the $\gamma$-ray
strength function without assuming any models for these functions.

The structure of the manuscript is as follows: Section 2 describes briefly the experimental set-up and the Oslo method,
and previously measured level densities will be discussed in Sect.~3. Finally, concluding remarks and outlooks are
presented in Sect.~4.

\section{The Oslo Method}
\label{sec:1}

The Oslo method is based on a set of $\gamma$ spectra as a function of excitation energy. The
excitation of the nucleus is performed by light ion reactions, e.g. $(d, p\gamma)$, $(p,p'\gamma)$ and $(^3$He, $\alpha\gamma)$
where the energy of the charged ejectile determines the excitation energy.

\begin{figure}[]
\begin{center}
\includegraphics[width=\columnwidth]{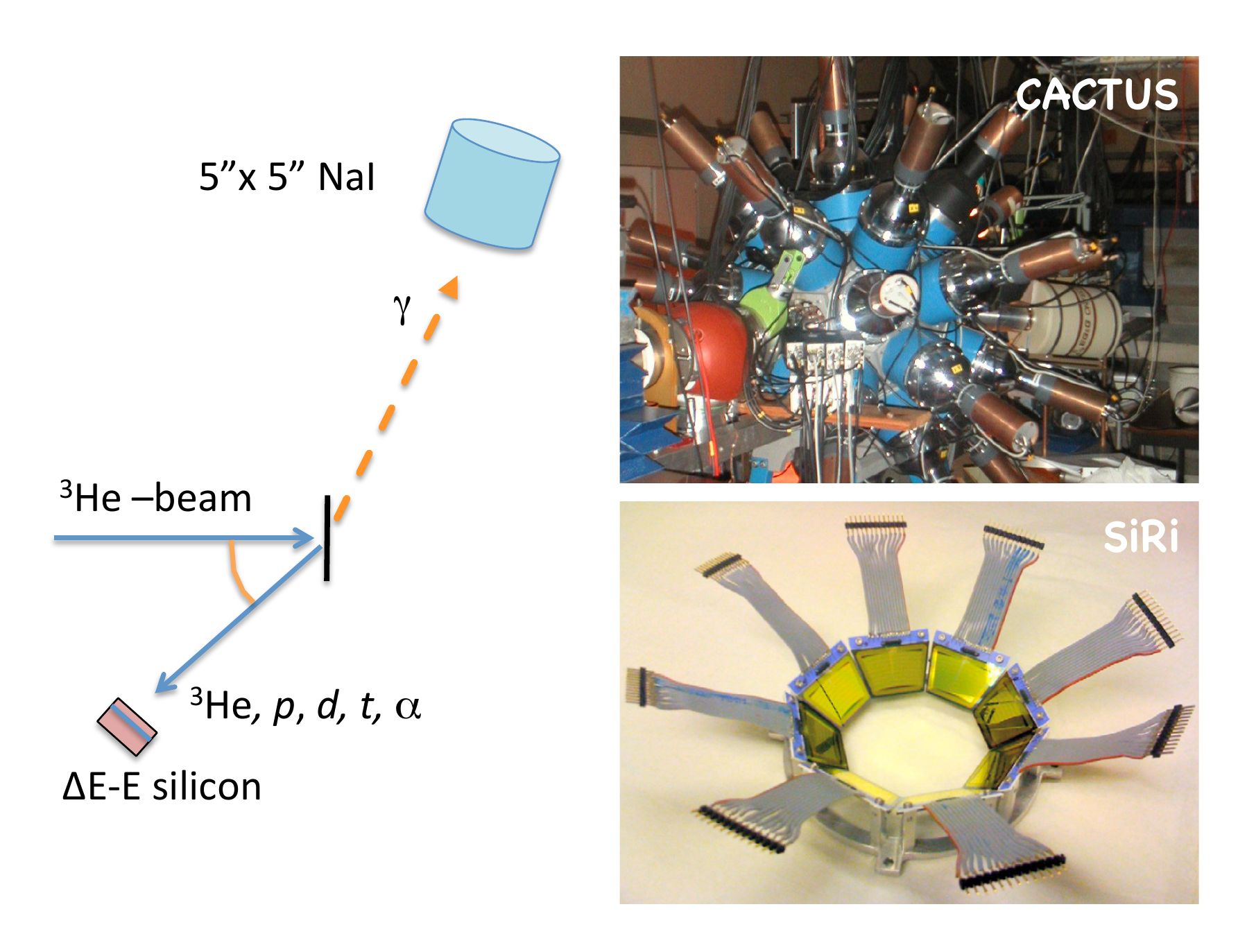}
\caption{(Color online) Typical particle-$\gamma$ coincidence set-up for the Oslo method.
The 64 silicon particle telescopes of SiRi are placed in the vacuum chamber at the center of CACTUS. }
\label{fig:set-up}
\end{center}
\end{figure}

A schematic drawing of the set-up is shown in Fig.~\ref{fig:set-up}. A silicon particle detection system (SiRi)~\cite{siri}, which
consists of 64 telescopes, is used for the selection of a certain ejectile type and to determine its energies. The
front $\Delta E$ and back $E$ detectors have thicknesses of 130 $\mu$m and 1550 $\mu$m, respectively. SiRi
is usually placed in backward angles covering $\theta = 126^\circ$ to $140^\circ$
relative to the beam axis. Coincidences with $\gamma$ rays are performed with the CACTUS array~\cite{CACTUS},
consisting of 26 collimated $5" \times 5"$ NaI(Tl) detectors with a total efficiency of $14.1$\% at $E_\gamma = 1.33$~MeV.

\begin{figure}[]
\begin{center}
\includegraphics[width=0.9\columnwidth]{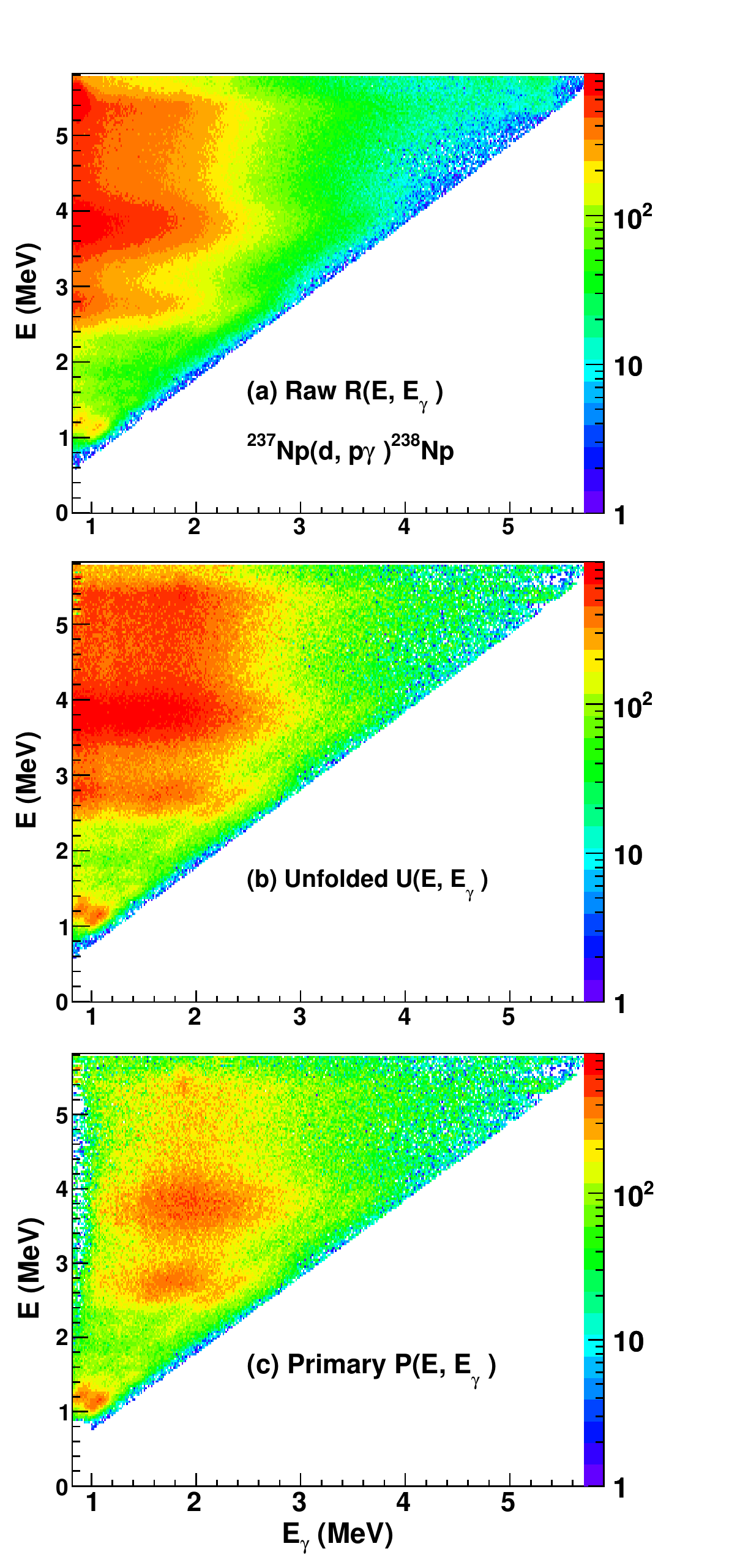}
\caption{(Color online) Initial excitation energy $E$ versus $\gamma$-ray
energy $E _{\gamma}$ from the $^{237}$Np$(d, p \gamma )^{238}$Np reaction~\cite{238Np}. The raw $\gamma$-ray spectra
(a) are first unfolded by the NaI response function (b) and then
the primary $\gamma$-ray spectra are extracted (c).}
\label{fig:matrices}
\end{center}
\end{figure}

Figure~\ref{fig:matrices} demonstrates the previously measured\\
$^{237}$Np$(d, p \gamma )^{238}$Np reaction~\cite{238Np} and
how one can proceed from the raw particle-$\gamma$ coincidences to the first generation or primary $\gamma$-ray spectra. The first step
is to sort the events into a raw particle-$\gamma$ matrix $R(E, E_{\gamma})$ with proper subtraction of
random coincidences. Then, for all initial excitation energies $E$, the $\gamma$ spectra are unfolded with the NaI response
functions giving the matrix $U(E, E_{\gamma})$~\cite{gutt1996}. The procedure is iterative and stops
when the folding ${\cal F}$ of the unfolded matrix equals the raw matrix within the statistical fluctuations, i.e.~when
${\cal F}(U)\approx R$.

\begin{figure}[]
\begin{center}
\includegraphics[width=0.8\columnwidth]{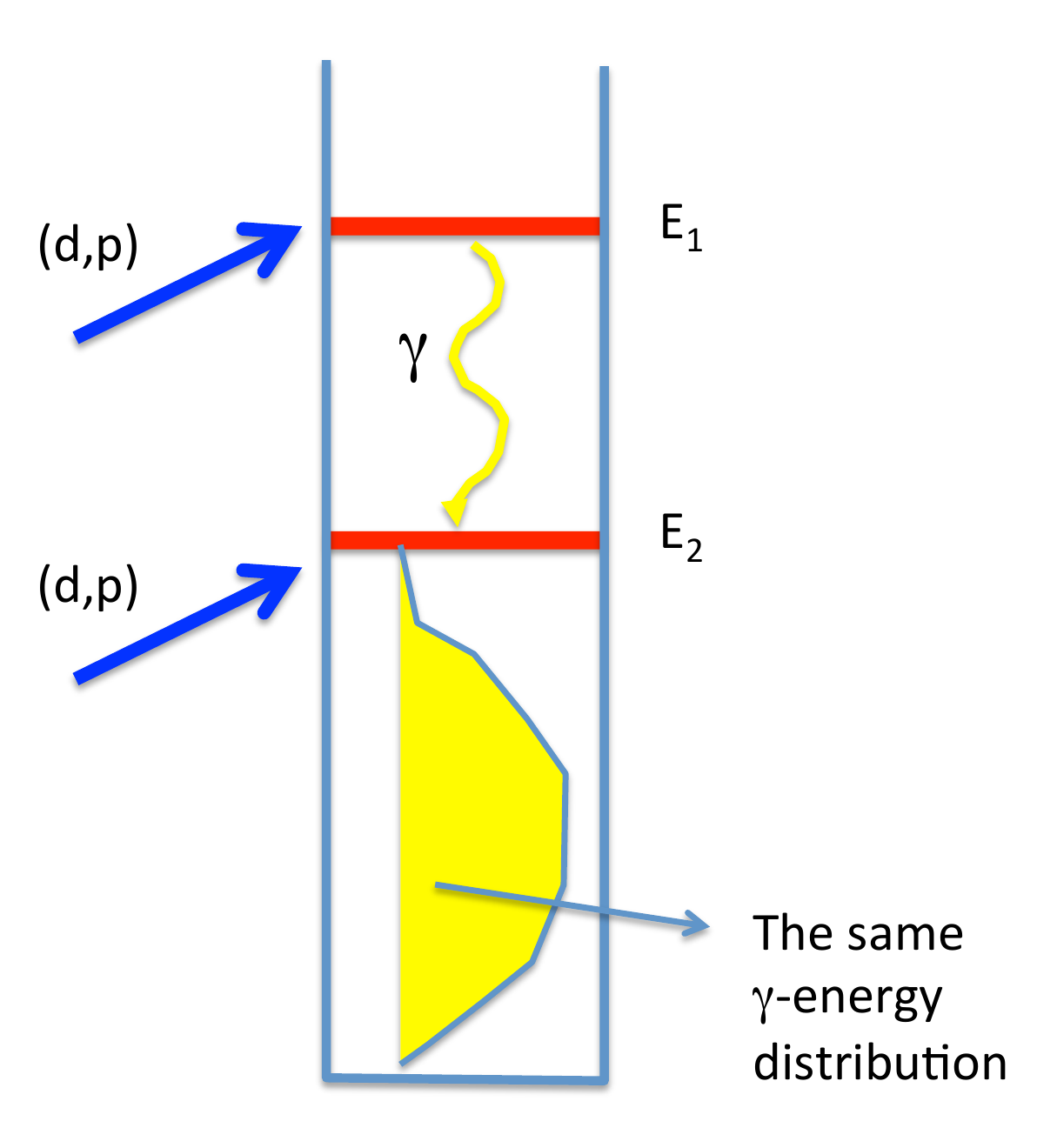}
\caption{(Color online) The assumption of the extraction of primary $\gamma$ spectra
is that we obtain the same energy distribution (yellow) whether the levels of the energy bin
$E_2$ are populated directly by e.g. the $(d, p)$ reaction or by $\gamma$ transitions from
excitation bin $E_1$.}
\label{fig:firstgen}
\end{center}
\end{figure}

The primary $\gamma$-ray spectra can be extracted from the unfolded total $\gamma$ spectra
$U$ of Fig.~\ref{fig:matrices} (b).
The primary $\gamma$ spectrum at an initial excitation energy $E$ is obtained by subtracting a weighted sum of
$U(E',E_{\gamma})$ spectra below excitation energy $E$:
\begin{equation}
P(E,E_{\gamma})=U(E,E_{\gamma}) - \sum_{E' < E}W(E,E')U(E',E_{\gamma}).
\end{equation}
The weighting coefficients $W(E, E')$ are determined in an iterative way described in Ref.~\cite{Gut87}.
After a few iterations, $W(E,E')$ converges to $P(E,E_{\gamma})$, where we have normalized
each $\gamma$ spectrum by $\sum_{E_{\gamma}}P(E,E_{\gamma})=1$.
This equality $P\approx W$ is exactly what is expected, namely that
the primary $\gamma$-ray spectrum
equals to the weighting function. The validity of the
procedure rests on the assumption that the $\gamma$-energy
distribution is the same whether the levels were populated directly by
the nuclear reaction or by $\gamma$ decay from
higher-lying states.
This is illustrated in Fig.~\ref{fig:firstgen}.

 The statistical part of this
landscape of probability, $P(E,E_{\gamma})$, is then assumed
to be described by the product of two vectors
\begin{equation}
P(E, E_{\gamma}) \propto   \rho(E-E_{\gamma}){\cal{T}}(E_{\gamma}) ,\
\label{eqn:rhoT}
\end{equation}
where the decay probability should be proportional to the
level density at the final energy $\rho(E-E_{\gamma})$ according
to Fermi's golden rule~\cite{dirac,fermi}. The decay is also proportional
to the $\gamma$-ray transmission coefficient ${\cal{T}}$, which
according to the Brink hypothesis~\cite{brink}, is independent of excitation energy;
only the transitional energy $E_{\gamma}$ plays a role.

\begin{figure*}[]
\begin{center}
\includegraphics[width=1.9\columnwidth]{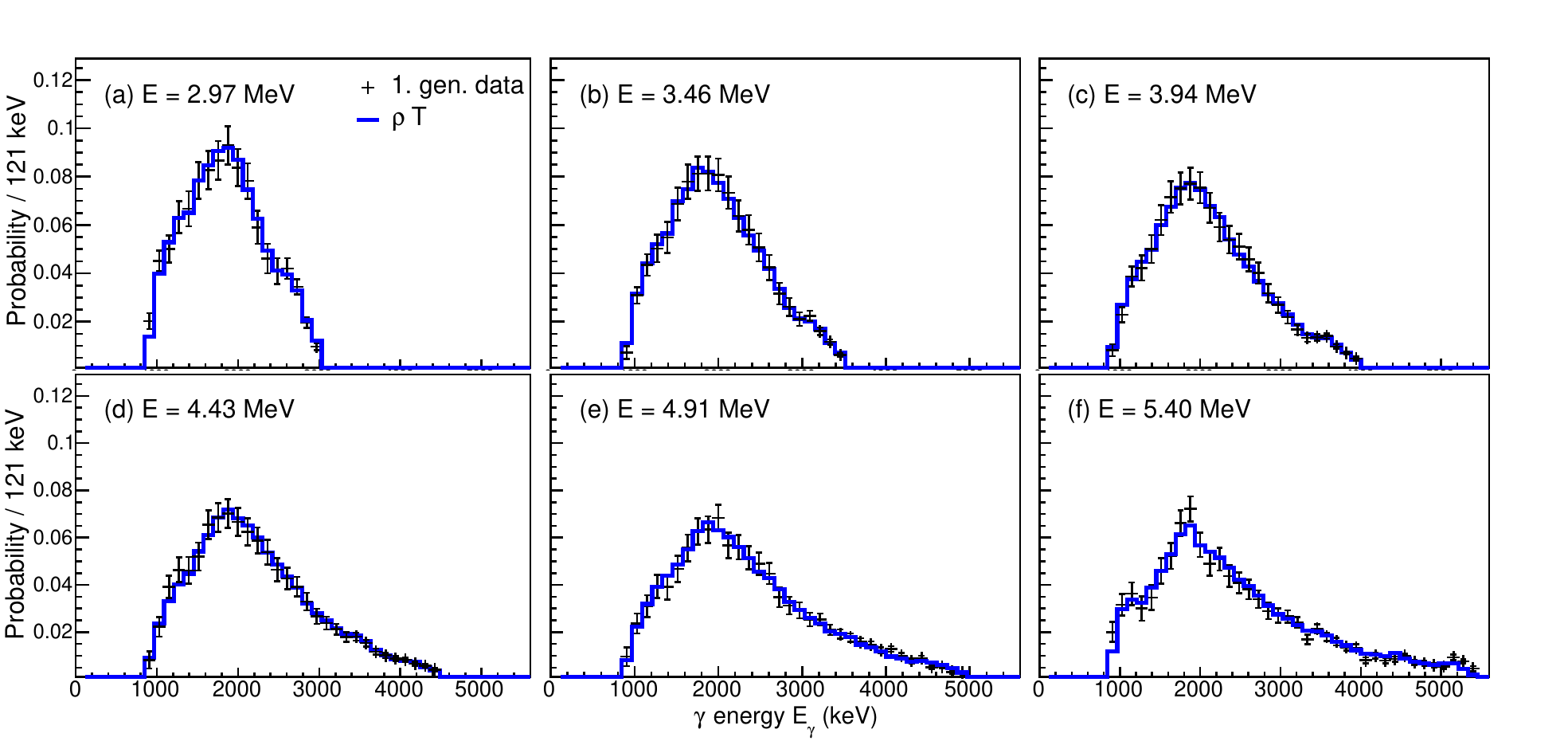}
\caption{(Color online) Primary $\gamma$-ray spectra from
the $^{237}$Np$(d, p \gamma )^{238}$Np reaction~\cite{238Np}
gated on various initial excitation energies $E$ (crosses).
The spectra are compared to the product of the level density and transmission coefficient functions
i.e.~$\rho(E-E_{\gamma}) {\mathcal{T}}(E_{\gamma})$ (blue lines).
The excitation energy gates are 121 keV broad.}
\label{fig:doesitwork}
\end{center}
\end{figure*}

\begin{figure}[h]
\begin{center}
\includegraphics[width=0.9\columnwidth]{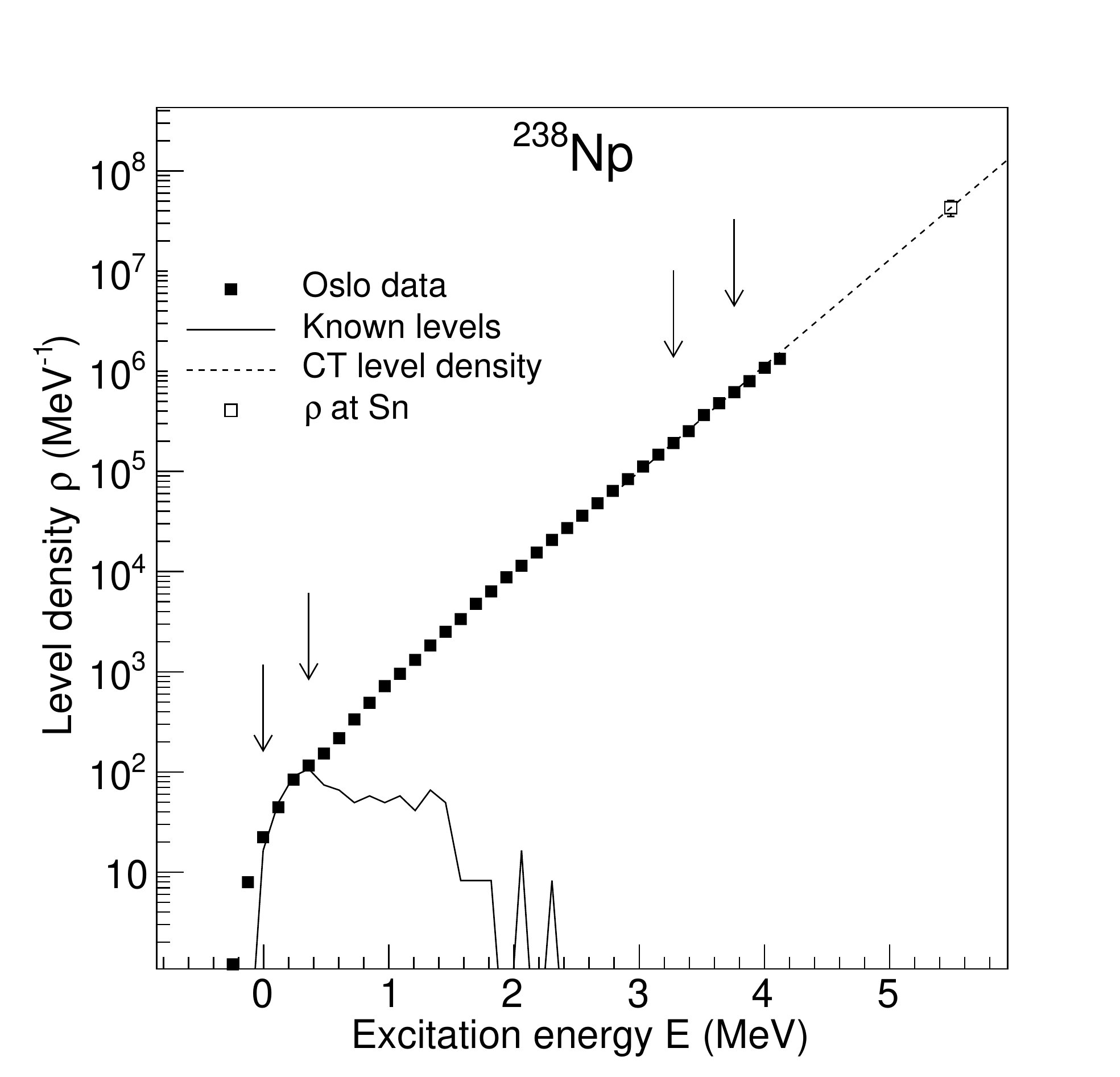}
\caption{Normalization procedure for the experimental level density.
The experimental data are normalized to known discrete levels
at low excitation energy and to the level density
extracted at $S_n$ from neutron capture resonance spacings $D_0$.
The two set of arrows indicate where the data are normalized by means of Eq.~(\ref{eq:array1}).
}
\label{fig:countingx}
\end{center}
\end{figure}

\begin{figure*}[tbh]
\begin{center}
\includegraphics[width=2\columnwidth]{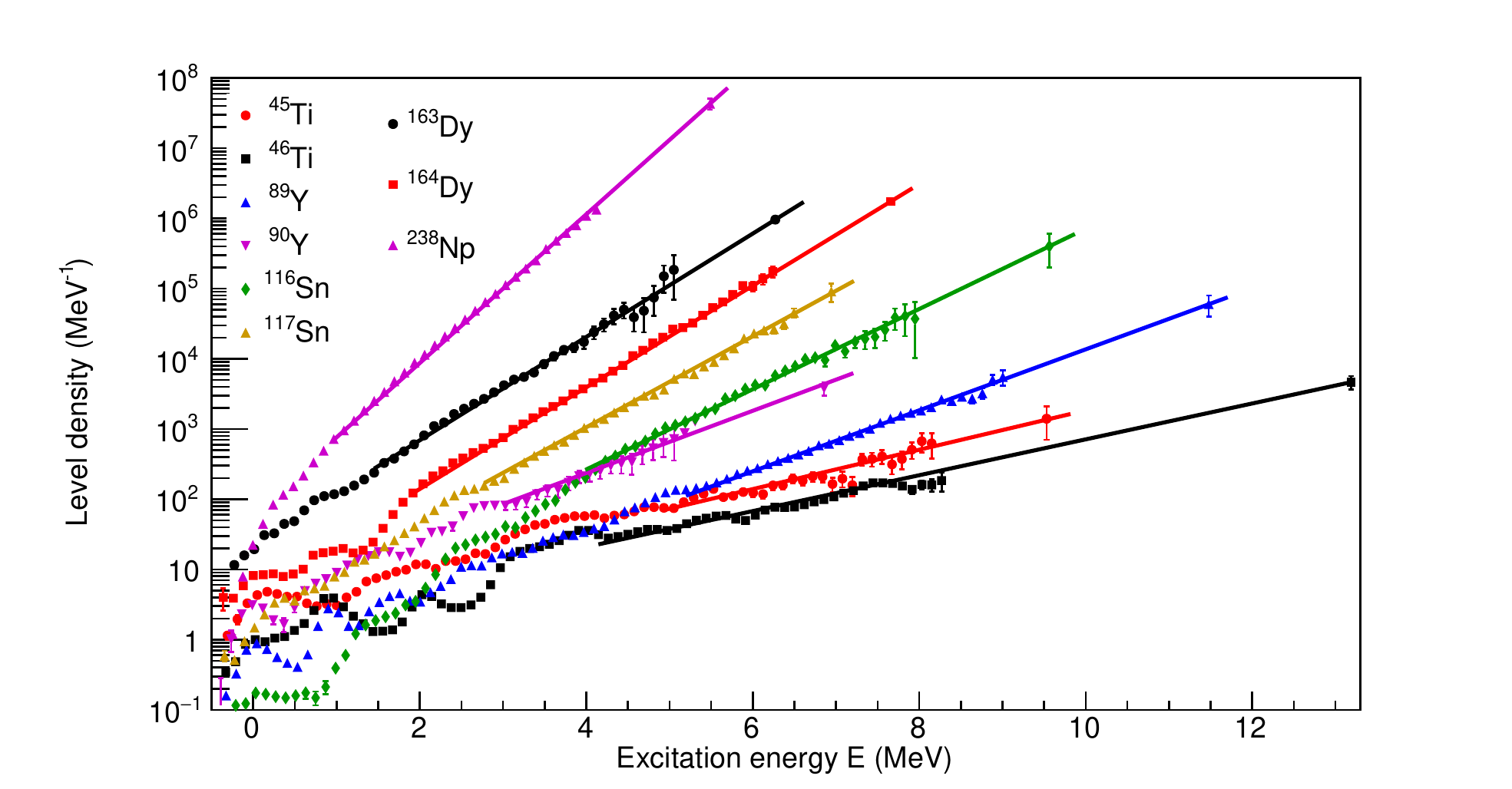}
\caption{(Color online) Experimental level densities obtained with the Oslo method.
For references, see Table~\ref{tab:parameters}.
}
\label{fig:rhotot_all}
\end{center}
\end{figure*}

The values of the vector elements of $\rho$ and  $\cal{T}$ are varied
freely to obtain a least-$\chi ^2$ fit to the experimental $P$ matrix.
Figure~\ref{fig:doesitwork} shows that the product of $\rho(E-E_{\gamma}){\cal{T}}(E_{\gamma})$ of Eq.~(\ref{eqn:rhoT})
fits very well the primary spectra of $^{238}$Np at the six excitation energy bins shown.
There are actually four times more spectra and the same quality of fit is obtained for all spectra.

However, the nice fit to the data does not mean that $\rho$ and  $\cal{T}$ are uniquely determined.
There are infinitely many functions that make identical
fits to the experimental $P$ matrix.
These $\tilde{\rho}$ and $\tilde{\cal{T}}$ functions can be generated by the transformations~\cite{Schiller00}
\begin{eqnarray}
\tilde{\rho}(E_i-E_\gamma)&=&A\exp[\alpha(E_i-E_\gamma)]\,\rho(E_i-E_\gamma),
\label{eq:array1}\\
\tilde{{\mathcal{T}}}(E_\gamma)&=&B\exp(\alpha E_\gamma){\mathcal{T}} (E_\gamma).
\label{eq:array2}
\end{eqnarray}
Thus, the normalization of the two functions requires information to fix
the parameters $A$, $\alpha$, and $B$, which is not available from our experiment.

For the level density, we use two anchor points to determine $A$ and $\alpha$.
The procedure is demonstrated for the case of $^{238}$Np in Fig.~\ref{fig:countingx}.
The level density is normalized to known discrete levels at low excitation energy.
At high excitation energy, we exploit the neutron resonance spacings $D_0$
at the neutron separation energy $S_n$. In order to translate the spacing
to level density, we  use the spin distribution~\cite{GC}
\begin{equation}
g(E,I) \simeq \frac{2I+1}{2\sigma^2}\exp\left[-(I+1/2)^2/2\sigma^2\right],
\label{eq:spindist}
\end{equation}
where $E$ is excitation energy and $I$ is spin. Here, the spin cut-off parameter
$\sigma$ is rather uncertain, and may introduce an uncertainty of a factor of 2 in $\rho(S_n)$.
For cases where $D_0$ is unknown, one has to use systematics to estimate $\rho(S_n)$.
Since the $\gamma$ strength is not discussed in this work, the
determination of the $B$ parameter is irrelevant.
Further description and tests of the Oslo method are given in Refs.~\cite{Schiller00,Lars11}.

\section{Experimental level densities}
\label{sec:2}

\begin{table*}[]
\begin{center}
\caption{Parameters to describe the level density.}
\begin{tabular}{l|cccc|r}
\hline
\hline
Reaction                 &$S_n$ &  $T_{\rm CT}$   & $E_0$  &  $\rho(S_n)$      & Refs.       \\
&(MeV) & (MeV)  & (MeV)  & ($10^6$MeV$^{-1}$)&             \\
\hline
$(p, p')^{45}$Ti      & 9.530& 1.55   & -2.33 & 0.0014(7)          &\cite{45Ti}  \\
$(d, p)^{46}$Ti       &13.189& 1.70   & -2.07 & 0.0047(10)         &\cite{46Ti}  \\
$(p, p')^{89}$Y       &11.478& 1.00   &  0.48 & 0.060(20)          &\cite{89Y}   \\
$(d, p)^{90}$Y        & 6.857& 1.00   & -1.38 & 0.0038(8)          &\cite{89Y}   \\
($^3$He,$\alpha)^{116}$Sn& 9.563& 0.76   & -0.03 & 0.40(20)           &\cite{117SnPRL,116117Snnld,116117Sn,121122Sn} \\
($^3$He,$^3$He$'$)$^{117}$Sn & 6.944& 0.67 & -0.40 & 0.091(27)          &\cite{117SnPRL,116117Snnld,116117Sn,121122Sn} \\
($^3$He,$\alpha$)$^{163}$Dy& 6.271& 0.59 & -1.55 & 0.96(12)           &\cite{nldDy} \\
($^3$He,$^3$He$'$)$^{164}$Dy & 7.658& 0.60 & -0.66 & 1.74(21)           &\cite{nldDy} \\
($d, p)^{238}$Np         & 5.488& 0.41   & -1.35 & 43.0(78)           &\cite{238Np} \\
\hline
\hline
\end{tabular}
\label{tab:parameters}
\end{center}
\end{table*}

The Oslo group has now published more than 70 level densities.
Some of these are for one and the same nucleus, but populated with different reactions.
The full compilation of level densities (and $\gamma$-ray strength functions)
is found in Ref.~\cite{compilation}. We will here focus on
some selected nuclei in order to convey to the reader a general picture
of how the level density behaves
as function of mass region and excitation energy. The references and key
numbers are given in Table~\ref{tab:parameters}.

There is a tremendous span in level densities when going from mass
around $A \approx 40$ to 240. Figure~\ref{fig:rhotot_all} shows that the
level density at excitation energy $E \approx 6$~MeV for $^{238}$Np
is almost ten million times higher than for $^{46}$Ti. It seems obvious
that more nucleons $A$ produce  higher level density, but this conclusion is
not strictly true.

Indeed, we see that the level density for $^{90}$Y is larger than for $^{89}$Y.
Also $^{117}$Sn has more levels than $^{116}$Sn.
However, $^{45}$Ti has more levels than $^{46}$Ti and the
same is true for $^{163}$Dy versus $^{164}$Dy. The pieces fall into place
if pairing is taken into account: For neighboring isotopes, the nucleus
with either one or more unpaired protons or neutrons displays  $5 - 7$ times more levels.
An increase in $A$ only gives negligible effects between neighboring nuclei,
as e.g.~found for the even-even $^{160,162,164}$Dy isotopes~\cite{luciano2014}.

The number
of quasi-particles is the basis building blocks for the level density.
However, the number of active particles making the level density is not that
different for $^{238}$Np and $^{46}$Ti. The odd-odd $^{238}$Np has about 6 quasi-particles at $E \approx 6$~MeV,
while $^{46}$Ti has about 2 quasi-particles.
The second factor needed to understand the general behaviour of the level density is the density of
available single-particle orbitals in
the vicinity of the Fermi level where the quasi-particles can play around.
In summary, the large span in level density
between $A = 40$ and 240 is a combined result of the number of quasi-particles and the density of
near lying single-particle orbitals.

Another striking feature of Fig.~\ref{fig:rhotot_all} is that,
above a certain excitation energy $E$, the level densities are linear
in a log-plot, which corresponds to an exponential function.
The $^{45}$Ti has an exponential behavior for $E > 4$~MeV,
whereas $^{238}$Np has it for $E > 1$~MeV.
Generally, the exponential behavior sets in when the first nucleon pairs
are broken, i.e. $E > 2 \Delta$. With a pairing gap of
$\Delta \approx 12A^{-1/2}$~MeV, we find for the two cases $E= 3.6$ and $1.6$ MeV, respectively.
The level density is not only dependent on the atomic mass number $A$,
but also on the number of unpaired nucleons around the ground state.
As an example, the exponential behavior is delayed for even-even nuclei compared to odd-even nuclei.
This is e.g. apparent for $^{116,117}$Sn and $^{163,164}$Dy.

The interpretation of the pure exponential growth above $E > 2 \Delta$ has recently been discussed in Ref.~\cite{luciano2014}.
For the micro-canonical ensemble,  the entropy is related to the level density by
\begin{equation}
S(E) \approx \ln \rho(E),
\end{equation}
with
\begin{equation}
T(E)=[\partial S(E)/\partial E]^{-1}.
\label{eqn:t}
\end{equation}
Thus, the temperature is constant when $\ln \rho(E)$ is linear in $E$. The common
expression for the constant-temperature level density is~\cite{Ericson}
\begin{equation}
\rho_{\rm CT}(E)=\frac{1}{T_{\rm CT}}\exp{\frac{E-E_0}{T_{\rm CT}}},
\label{eq:ct}
\end{equation}
where $T_{\rm CT}$ is determined by the slope of $\ln \rho(E)$.
This is the key characteristic of a first-order phase transition.
Our interpretation of the data is that the energy goes into breaking
Cooper pairs and thus the temperature remains constant. The
obvious analogy is the melting of ice to water at constant temperature.
The fit parameters $T_{\rm CT}$ and $E_0$ for
the nuclei shown in Fig.~\ref{fig:rhotot_all} are listed in Table~\ref{tab:parameters}.

\begin{figure}[b]
\begin{center}
\includegraphics[width=1\columnwidth]{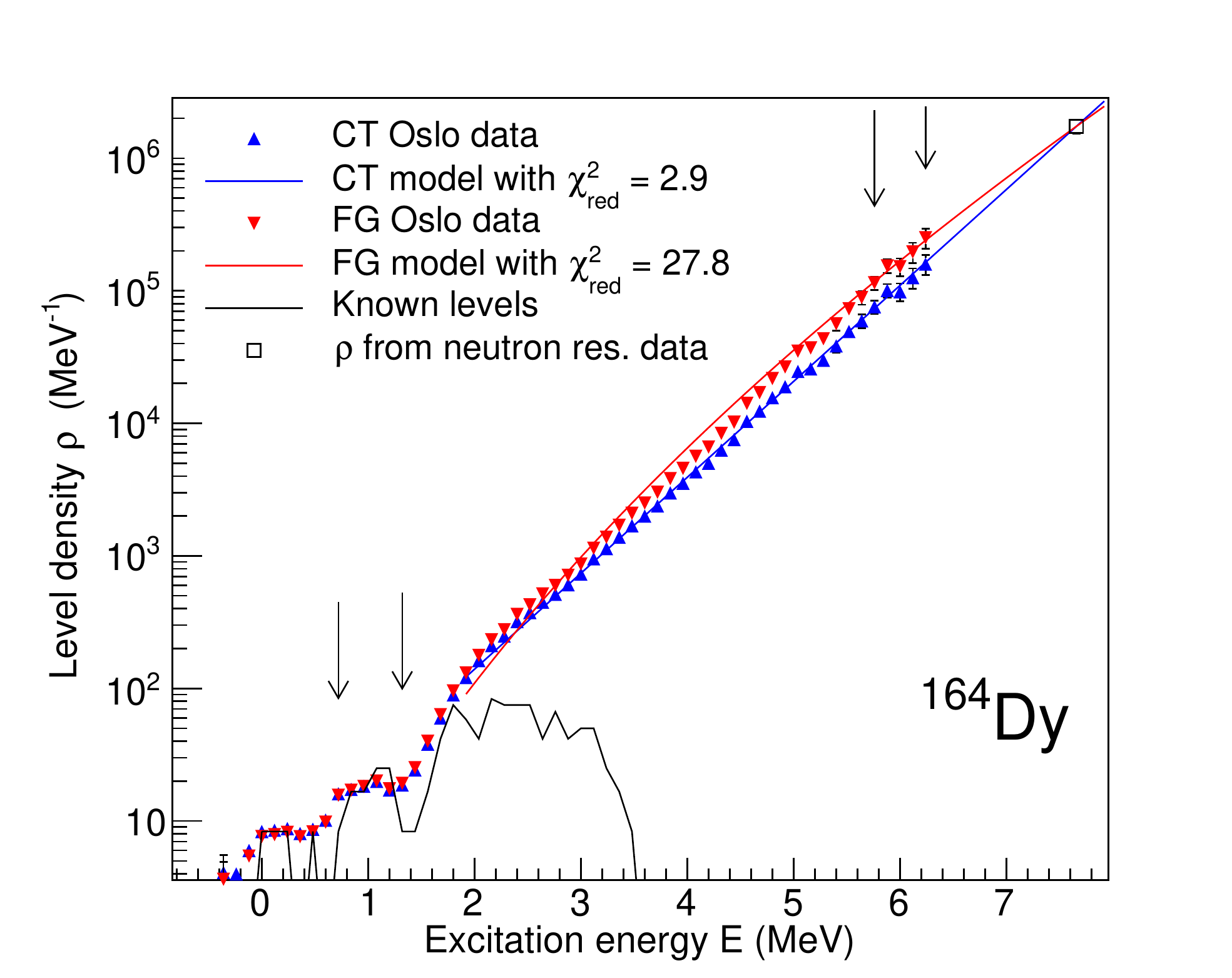}
\caption{(Color online) Experimental level densities of $^{164}$Dy compared
with the constant-temperature (CT) and the Fermi gas (FG) level density formulas.
The reduced least-$\chi ^2$ fit is best for the CT model.}
\label{fig:ctfg}
\end{center}
\end{figure}

Figure \ref{fig:ctfg} shows that the constant-temperature level density model fits
the data of $^{164}$Dy much better than the Fermi gas model. The reduced $\chi ^2$ fit value
evaluated between excitation energies $E= 1.9$ and $6.2$ MeV is about a factor of 10 better
for the CT model. One should of course mention that the CT model is not a perfect model.
There are some oscillations in the experimental log-plot mainly due to the onset of the
two and four quasi-particle regimes.

Recommendations are available for the Fermi-gas spin-cut off
parameter $\sigma (A,E)$ \cite{Egidy2009}. However, for the CT-model there are
no clear recommendations. According to the original expression of Ericson~\cite{Ericson}:
\begin{equation}
\sigma^2= \frac{\Theta T}{\hbar^2},
\end{equation}
we see that for the CT model, only the moment of inertia $\Theta$ will depend on $E$,
since $T=T_{CT}$ is a constant. There are indications~\cite{Alhassid2007} that the rigid moment
of inertia is reached at the neutron separation energy, i.e.~$\Theta(E=S_n)=\Theta_{\rm rigid}$.
However, how to parameterize $\sigma$ to lower excitation energies is still uncertain.
In addition to the uncertain spin distribution, the parity distribution is important for
lighter nuclei and nuclei close to magic numbers. There are only few experimental
data available in order to pin down the spin and parity distributions and this
represents perhaps the most important issue for future research on nuclear level densities.

The pair breaking process will continue at least up to the binding energy, but probably
to much higher excitations. In several evaporation experiments the
constant temperature behavior has been observed up to $E \approx 15$~MeV \cite{Voinov2009,Byun2014}.
However, at the higher excitation energies the temperature will start increasing with
excitation energies, and the Fermi gas description~\cite{bethe36,capote2009} will to an increasing degree describe the data.

\section{Concluding remarks and outlook}

The present work describes the Oslo method and results using light-ion reactions on stable targets.
The method is demonstrated by means of the $^{237}$Np$(d, p \gamma )^{238}$Np reaction, which provides almost ideal
conditions for the Oslo method.   One important aim of these investigations is to obtain more
predictive power for extrapolation of the level densities to nuclei outside the valley of $\beta$-stability.

From light ($^{45}$Ti) to the heavy elements ($^{238}$Np) we find certain common features of the level densities.
It turns out that the pairing force is responsible for the mechanism behind these features.
The level density above the excitation energy of 2$\Delta$ behaves as an exponential curve
characterized by the nuclear temperature. Furthermore, neighboring nuclei have the same
temperature and their respective level densities appear parallel in a log plot.
Odd-mass nuclei exhibit about $5 - 7$ times more levels than their even-even partners.
There are indications that each unpaired nucleon carry approximately an entropy of $1.6 - 1.9 k_B$,
corresponding to the multiplicative factor above.

In the field of experimental level densities, there are several skeletons in the closet.
In particular, it is crucial to get more information on the true spin and parity distributions
as function of excitation energy and mass. The few experimental data on these quantities
also have implications for applications when e.g. calculating reaction rates.
Today, such information is often taken from theoretical models
without verifications based on experimental observations.

In the energy regime from the ground state up to the energy where the breaking
of the first Cooper pairs set in, the level density depends strongly on the
specific nuclear structure properties of the nucleus studied. For even-even nuclei,
the level density is determined by the collective modes like rotation and vibrations.
For the odd-mass nuclei the unpaired valence nucleon already carries a quasi-particle
entropy that adds to these collective degrees of freedom.

A great challence is to obtain data also for nuclei far away from the $\beta$-stability line,
nuclei that are essential for nucleo-synthesis networks and other applications.
Since it is a rather uncertain approach to extrapolate results from more stable nuclei
to these nuclei, novel experimental techniques are needed.
The Oslo group has already initiated experiments to measure level densities for
short-lived nuclei. The so-called $\beta$-Oslo method is based on the population of
rare nuclei by means of $\beta$ decay. The first experiment, which was performed at
the National Superconducting Cyclotron Laboratory at Michigan State University (MSU)
using the SuN detector has already been published~\cite{spyrou}, and several new experiments
have been carried out. Also inverse kinematic experiments with radioactive beams are planned at HIE-ISOLDE.

The constant-temperature level density behavior is expected also for nuclei far
from the $\beta$-stability line. However, when the particle separation energies
approaches the paring energy of $E = 2\Delta$, the individual low-lying discrete levels will
play an increasing role. Furthest away from the valley of stability, discrete spectroscopy of
low-lying levels will probably be the way to proceed. In such scenarios the concept
of level density will lose its importance and be replaced by discrete spectroscopy.

\label{sec:3}

\end{document}